\documentclass[twocolumn,showpacs,epsfig,prd,preprintnumbers,amsmath,amssymb]{revtex4}
\usepackage{graphicx}
\usepackage{epsfig}
\usepackage{epsf}
\usepackage{dcolumn}
\usepackage{bm}

\newcommand{\ccbar}{c\bar c}

\def \qqbar {q\bar q}
\def \jp {J/\psi}

\newcommand{\ppbar}{p\bar p}

\newcommand{\chis}{\chi^2}
\newcommand{\psip}{\psi(2S)}

\def \kk {K^+K^-}
\def \pp {\pi^+\pi^-}

\def \piz {\pi^0}

\def \hd {\textrm{hadrons}}
\def \bg {background~}
\def \g {\gamma}

\def \cfm {$\chis$ fit method}

\begin{document}

\title{Extract Signals by Fitting $\chi^2$ Distribution of the Kinematic Fit}

\thanks{This work is partly supported by the National Natural Science
Foundation of China under Grant Nos. 10375074 and 10491303 and the
100 Talents Program of CAS under Contracts No. U-25}

\author{Ping Rong-Gang, Mo Xiao-Hu, Du Shu-Xian, Liu
Jing,\\ Ma Lian-Liang,  Shen Cheng-Ping, Wang Zhi-Yong and Yuan
Chang-Zheng}

\affiliation{Institute of High Energy Physics Chinese Academy of
Sciences\\ P.O. Box 918(1), Beijing 100049}
\thanks{correspondence author: pingrg@mail.ihep.ac.cn, Tel: 010-88233823}

\begin{abstract}

In measuring the $\psip$ radiative decays at BESII, contribution of
the background is serious in most of the final states. To extract
the number of signal events, a fit to the $\chi^2$ distribution of
kinematic fit using signal and background components is proposed.
Extensive Monte-Carlo simulations (MCS) are performed, and the
results show that the shape of $\chi^2$ distribution of the signal
channel looks different from those of the background channels, thus
it provides us by fitting the $\chi^2$ distribution of the data to
extract the number of signal events. An input-output test is
performed using MCS, and the uncertainty of the fit method is found
to be less than 2\%.

\end{abstract}



\maketitle

\vspace{1.5cm}

\section{INTRODUCTION}

The radiative decays of $\jp$ and $\psi'$, due to the direct
coupling of outgoing photon to the charm quark, could be an
excellent laboratory to study the hadron spectroscopy and search for
exotic states. At the leading order of pQCD picture, both the $\jp$
and $\psi$ radiative decays can be described by the $\ccbar$ quark
annihilation with the emission of one photon and two gluons. There
could be a possibility that the two gluons might form a glueball, or
one gluon combines with a $\qqbar$ pair to form a hybrid state, or
the two gluons couple to a multiquark state. Compared with $\jp$
decays, few measurements of $\psip$ radiative decays have been
reported~\cite{pdg2004}. BESII detector has accumulated 14 million
$\psip$ events, and it offers an opportunity to study a number of
$\psip$ radiative decays.

Experimentally, an event is reconstructed by using the detector
information on charged tracks and photons. For removing background
events, the constraints of the energy and momentum conservation for
the final state of the signal channel are imposed on the fitting to
the 4-component momentum (4C) reconstructed by the detector, and the
background events can be suppressed by using information of the
distribution of the $\chi^2$ of the fit. With the help of the
Monte-Carlo simulation, a reasonable requirement on $\chis$ can be
obtained by comparing the $\chis$ distributions between the signal
and the background channels. However, the $\chis$ cut method fails
badly sometimes if the $\chis$ distribution of the background
channel overlaps largely with the signal channels. In this paper, we
try to fit the $\chis$ distribution to extract the number of signal
events in measuring the exclusive decays of $\psip\to\gamma +\hd$ at
BESII detector, which includes 2, 4 and 6 prongs (stable
hadron---pion,kaon,proton or anti-proton).

\section{Fit to the $\chi^2$ distribution}

Monte Carlo simulation shows that backgrounds of the radiative decay
of $\psip\to\gamma+\textrm{hadrons}$ are largely coming from
multi-photon decay channels, namely, the dominant background is
$\psip\to\pi^0+\hd$, and some contaminations are from
$\psip\to\gamma\piz+\hd$, along with other possible backgrounds. In
principle, their $\chis$ distributions of the kinematic fit can be
obtained with MCS. Since these types of backgrounds are due to
missing photons, their distributions of the $\chis$ shapes are
different from each other. This property allows us to express the
observed $\chis$ distributions of the data sample by that of the
signal channel and the background channels, $i.e.$
\begin{equation}\label{csd}
\chis_{obs}=w_s\chis_{sig}+\sum_{w_{b_i}}w_{b_i}\chis_{bg},
\end{equation}
where $w_s$ and $w_{b_i}$ are the weight of the signal and the
background decays, respectively. If the background channels are
completely known, then the signal events can be extracted reliably
by fitting to the data with Equation~(\ref{csd}).

\section{Monte Carlo Simulation}

\subsection{Background}

To make the measurement of $\psi'$ radiative decays reliable, the
\bg should be fully studied. With the help of SIMBES
system~\cite{simbes}, extensive simulations have been performed.
Table~\ref{mcbg} summarizes the possible background contributions to
signal mode $\psip\to\gamma +\hd$. Generally speaking, the dominant
backgrounds come from multi-photon decays, i.e. $\psip\to
n\gamma+\hd~(n\geq 2)$. With more missing photons in \bg decays, the
contamination to the signal decay becomes lower and lower.

\begin{table}[htbp]
\parbox{0.45\textwidth}{\caption{The backgrounds for the radiative decay
$\psip\to\gamma+\hd$, where the normalized \bg events are based on
14M $\psi(2S)$ events and the branching fractions given in PDG table
with requirement $m_{\hd}<3.0$~GeV. \label{mcbg}}}
\begin{center}
\begin{tabular}{l|lc}\hline\hline
Signal mode&Background modes& $N_{bg}$(normalized)\\
$\psip\to$&$\psip\to$&\\\hline
$\gamma p\bar p$&$\pi^0 p\bar p$&$141.5$\\
&$\gamma\pi^0 p\bar p$&$8.8$\\\hline
$\g\kk\pp$&$\piz\kk\pp$&$98$\\
&$\g\piz\kk\pp$&$16$\\\hline
$\gamma 2(\pi^+\pi^-)$&$\pi^0 2(\pi^+\pi^-)$&805.0\\
&$\pi^+\pi^-\jp,\jp\to 3\pi$&33.4\\\hline
$\gamma K^+K^-2(\pi^+\pi^-)$&$\pi^0 K^+K^-2(\pi^+\pi^-)$&6.0\\
&$\gamma\pi^0 K^+K^-2(\pi^+\pi^-)$&6.1\\\hline\hline
\end{tabular}
\end{center}
\end{table}

\subsection{$\chis$ distribution}

\begin{figure}[htbp]
\begin{center}
\vspace*{-0.5cm}
\begin{center}
\hspace*{-0.cm} \epsfysize=8cm \epsffile{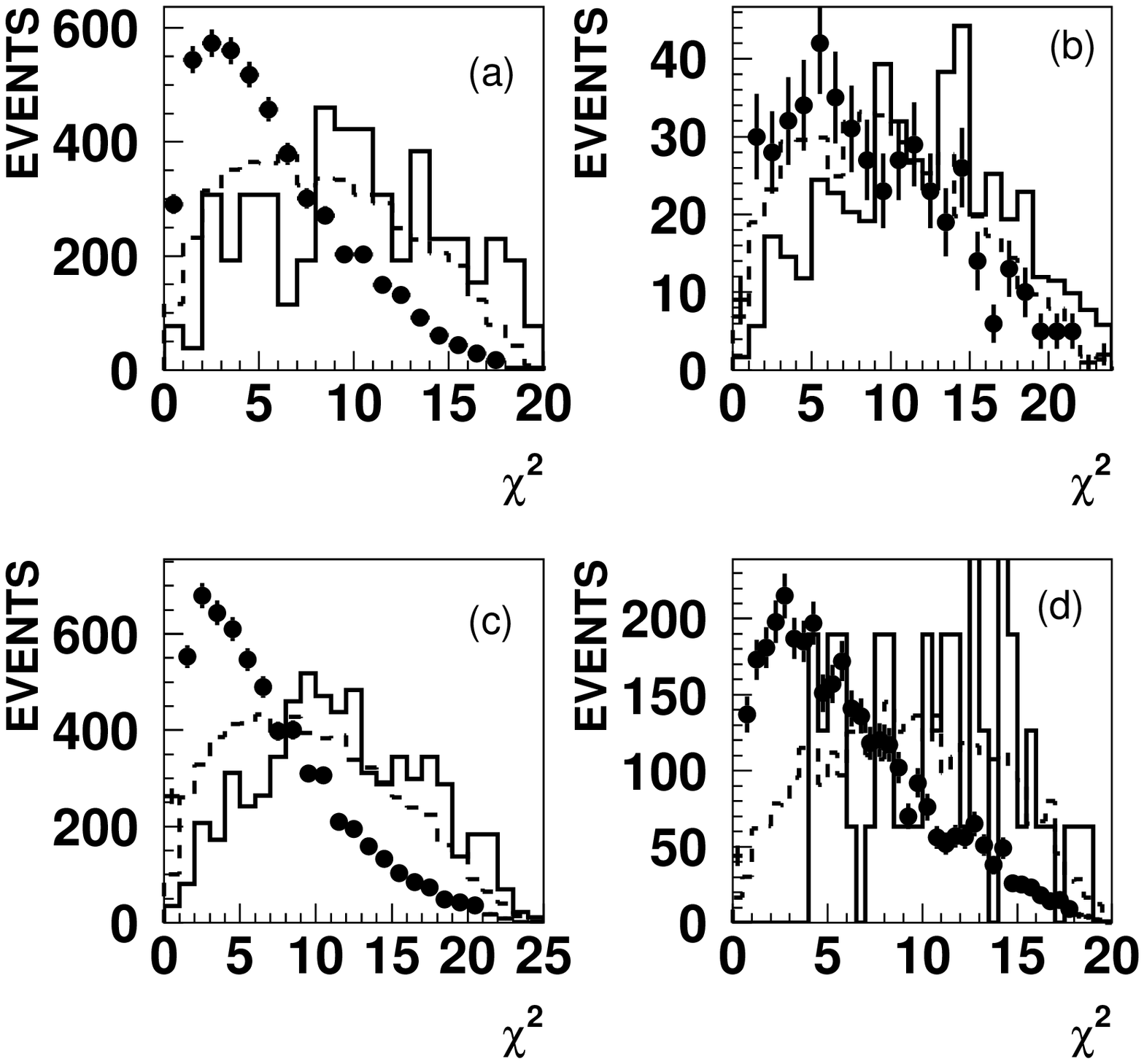}
\end{center}
\vspace*{0cm}\parbox{0.45\textwidth}{\caption{The chisqured
distribution of MC samples for $\psip\to\g+\hd$, where points with
error bars show the $\chis$ shapes for $\hd=p\bar
p~(a),~\kk\pp~(b),~2(\pp)~(c)$ and $\kk2(\pp)~(d)$, and the dashed
histograms correspond to $\psip\to \piz+\hd$ and the solid line
histograms to $\psip\to\g\piz+\hd$, respectively. \label{chisdis}}}
\end{center}
\end{figure}

Figure~\ref{chisdis} shows the shapes of $\chis$ distributions for
$\psip\to\g+\hd$ after data selection. In plots the points with
error bars are the signal decays into $\g\ppbar~(a)$,
$\g\kk\pp~(b)$, $\g2(\pp)~(c)$ and $\g\kk2(\pp)~(d)$, respectively.
The dashed histograms and the solid line histograms correspond to
the distributions of the background channels $\psip\to\piz+\hd$ and
$\g\piz+\hd$, respectively. For comparison of their shapes, the
background events are respectively normalized to signal events.
Obviously it can be seen that the shapes of the signal $\chis$
distributions are different from those of the background modes. This
could be understandable due to the fact that the expectation of the
$\chis$ value of the kinematic fit becomes larger to the backgrounds
originated from one or more photons missing.

\subsection{Input-output check of $\chis$ fit method}

Since the dominant backgrounds of $\psip$ radiative decays originate
from the photon missing decays, and their shapes are different from
each other, therefore, fitting the $\chi^2$ distribution of the data
with Equation (\ref{csd}) provides us a tool to extract the signal
events. We employ MCS to make an input-output check and estimate the
uncertainty of the $\chis$ fit method.

\begin{figure}[htbp]
\begin{center}
\vspace*{-0.5cm}
\begin{center}
\hspace*{-0.cm} \epsfysize=7.5cm \epsffile{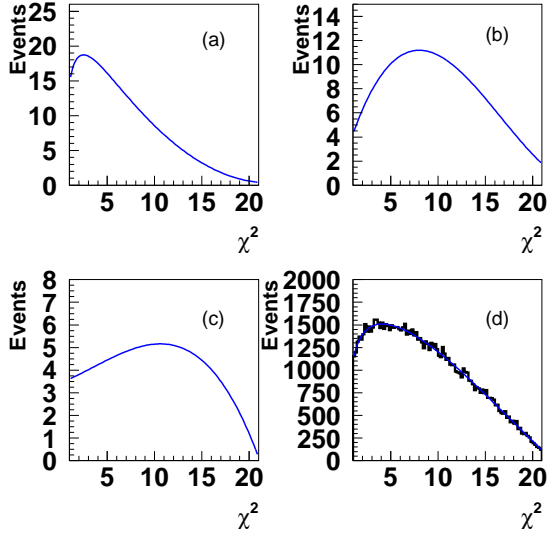}
\end{center}
\vspace*{0cm}\parbox{0.45\textwidth}{\caption{The $\chis$
distribution for $\psip$ decays into the signal mode $\g\kk\pp$~(a),
and the dominant background $\piz\kk\pp$ (b) and other unknown
background (c). In plot (d) the histogram corresponds to the $\chis$
distribution of the sample generated with $\chis$ shapes in (a), (b)
and (c) and the solid line to the fitted results with these $\chis$
components. \label{chischeck}}}
\end{center}
\end{figure}

Figure~\ref{chischeck} shows the $\chis$ distributions of the MC
sample for $\psip$ decays into $\g\kk\pp$~(a), $\piz\kk\pp$~(b) and
other unknown backgrounds by using 14M $\psip$ Lund-Charm MC sample.
In order to make an input-output check of the $\chis$ fit method, a
MC sample is generated with the $\chis$ shapes given in (a), (b) and
(c) with an input number $N_{\textrm{sig}}^{\textrm{in}}$,
$N_{\textrm{bg}}^{\textrm{in}}$ and $N_{\textrm{un}}^{\textrm{in}}$,
respectively. Then we extract the numbers of events of these modes
by the $\chis$ fit method. The total uncertainty includes errors
from the $\chis$ fit method and the statistics, {\it i.e.}
 \begin{equation}\label{}
\sigma_{\chis}=\sqrt{\left({N_{\textrm{sig}}^{\textrm{out}}-
N_{\textrm{sig}}^{\textrm{in}}\over N_{\textrm{sig}}^{\textrm{in}}
}\right)^2+\sigma_{\textrm{st}}^2},
 \end{equation}
where $N_{\textrm{sig}}^{\textrm{out}}$ is the number of the signal
events obtained by the \cfm, and $\sigma_{\textrm{st}}$ is the
statistical uncertainty.  Due to the uncertainty of the unknown
background shape, we also consider the unknown background shape
changed between the shape as shown in Figures~\ref{chischeck}~(b)
and (c) parameterized by
$$\chis_{\textrm{un}}\to (1-x)\chis_{\textrm{un}}+x\chis_{\textrm{bg}},$$
where $\chis_{\textrm{un}}$ and $\chis_{\textrm{bg}}$ are the
$\chis$ distribution of the unknown background and the dominant
background, respectively. The $x$ value is taken between zero to
one. The MCS shows that if the $\chis$ shape of the unknown
background tends to that of the dominant background, the uncertainty
of extracting the number of the dominant background becomes large by
\cfm. However, this situation does not worsen the signal
uncertainty. The MCS with a large statistics shows that the
uncertainty of the \cfm ~is less than 2\%.

\section{SUMMARY}
Based on MCS, it is found that the backgrounds of the $\psip$ decays
into $\g+\hd$ dominantly originate from the missing photon decays
like $\psip\to n\g+\hd~ (n \ge 2)$. The $\chis$ distribution shapes
of backgrounds are distinctive from those of the signal decays. This
property allows us to extract the number of the signal events by the
fit to the $\chis$ distribution. The results of MCS study indicate
that the uncertainty of the extracted signal events from the method
is less than 2\%.

\begin{acknowledgments}
We would like to thank Prof. Zhu Yong-Sheng and Dr. Li Gang for
useful comments.
\end{acknowledgments}

\end{document}